\documentclass[fleqn,10pt]{SelfArx} 

\definecolor{color1}{RGB}{0,0,90} 
\definecolor{color2}{RGB}{0,20,20} 

\usepackage{hyperref} 
\hypersetup{hidelinks,colorlinks,breaklinks=true,urlcolor=color2,citecolor=color1,linkcolor=color1,bookmarksopen=false,pdftitle={Title},pdfauthor={Author},urlcolor=blue}

\usepackage{graphicx}
\usepackage{natbib}
\usepackage{amsmath}
\usepackage{multirow}
\usepackage{subfigure}
\usepackage{epstopdf}
\usepackage{eucal}

\newcommand{\pd}{\partial}
\newcommand{\Brem}{B_{\textrm{\scriptsize{rem}}}}
\newcommand{\Rce}{R_\textrm{c}}
\newcommand{\Rin}{R_\textrm{i}}
\newcommand{\Rou}{R_\textrm{o}}
\newcommand{\Ren}{R_\textrm{e}}
\newcommand{\Acon}[1]{\mathcal{A}^{#1}}
\newcommand{\Bcon}[1]{\mathcal{B}^{#1}}

\JournalInfo{Published in Journal of Magnetism and Magnetic Materials, Vol. 322 (1), 133–141, 2014} 
\Archive{\href{http://dx.doi.org/10.1016/j.jmmm.2009.08.044}{DOI: 10.1016/j.jmmm.2009.08.044}} 

\PaperTitle{Analysis of the magnetic field, force, and torque for two-dimensional Halbach cylinders} 

\Authors{R. Bj\o{}rk, C. R. H. Bahl, A. Smith and N. Pryds} 
\affiliation{\textit{Department of Energy Conversion and Storage, Technical University of Denmark - DTU, Frederiksborgvej 399, DK-4000 Roskilde, Denmark}} 
\affiliation{*\textbf{Corresponding author}: rabj@dtu.dk} 

\newcommand{\keywordname}{Keywords} 

\Abstract{The Halbach cylinder is a construction of permanent magnets used in applications such as nuclear magnetic resonance apparatus, accelerator magnets and magnetic cooling devices. In this paper the analytical
expression for the magnetic vector potential, magnetic flux density and magnetic field for a two dimensional Halbach cylinder are derived. The remanent flux density of a Halbach magnet is characterized by the integer $p$.
For a number of applications the force and torque between two concentric Halbach cylinders are important. These quantities are calculated and the force is shown to be zero except for the case where $p$ for the inner magnet
is one minus $p$ for the outer magnet. Also the force is shown never to be balancing. The torque is shown to be zero unless the inner magnet $p$ is equal to minus the outer magnet $p$. Thus there can never be a force and a
torque in the same system.}

\begin{document}

\flushbottom 

\maketitle 


\thispagestyle{empty} 

\section{Introduction}\label{Sec.Introduction}
The Halbach cylinder \cite{Mallinson_1973,Halbach_1980} (also known as a hole cylinder permanent magnet array (HCPMA)) is a hollow permanent magnet cylinder with a remanent flux density at any point that varies 
continuously as, in polar coordinates,
\begin{eqnarray}
B_{\mathrm{rem},r}    &=& B_{\mathrm{rem}}\; \textrm{cos}(p\phi) \nonumber\\
B_{\mathrm{rem},\phi} &=& B_{\mathrm{rem}}\; \textrm{sin}(p\phi)\;,\label{Eq.Halbach_magnetization}
\end{eqnarray}
where $B_{\mathrm{rem}}$ is the magnitude of the remanent flux density and $p$ is an integer. Subscript $r$ denotes the radial component of the remanence and subscript $\phi$ the tangential component. A positive value of 
$p$ produces a field that is directed into the cylinder bore, called an internal field, and a negative value produces a field that is directed outwards from the cylinder, called an external field.

\begin{figure}
\includegraphics[width=\columnwidth]{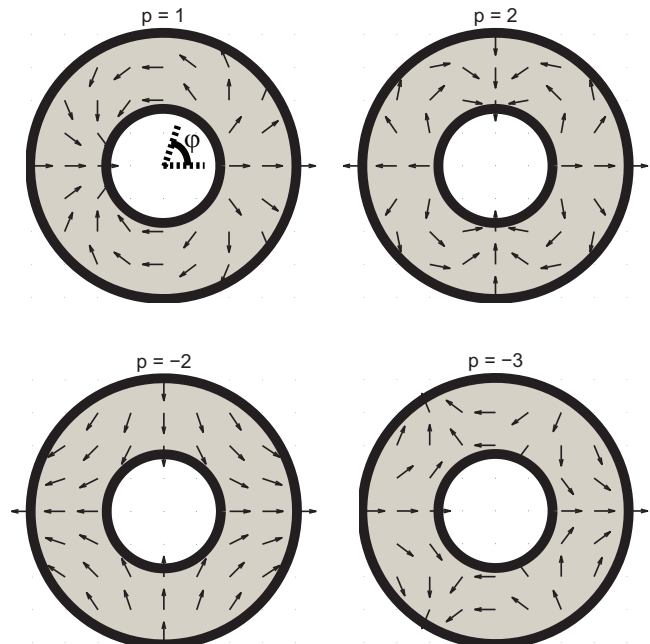}
      \caption{The remanence of a $p=1$, $p=2$, $p=-2$ and $p=-3$ Halbach cylinder. The angle $\phi$ from Eq. (\ref{Eq.Halbach_magnetization}) is also shown.}\label{Fig.Magnetization_drawing}
\end{figure}

A remanence as given in Eq. (\ref{Eq.Halbach_magnetization}) can, depending on the value of $p$, produce a completely shielded multipole field in the cylinder bore or a multipole field on the outside of the cylinder. In 
Fig. \ref{Fig.Magnetization_drawing}  Halbach cylinders with different values of $p$ are shown.

The Halbach cylinder has previously been used in a number of applications \cite{Zhu_2001, Coey_2002}, such as nuclear magnetic resonance (NMR) apparatus \cite{Appelt_2006}, accelerator magnets \cite{Lim_2005} and magnetic 
cooling devices \cite{Tura_2007}.

In these applications it is very important to accurately calculate the magnetic flux density generated by the Halbach cylinder. There exist several papers where the magnetic field and flux density for some parts of a 
Halbach cylinder are calculated \cite{Zhu_1993,Atallah_1997,Peng_2003,Xia_2004}, but a complete spatial calculation as well as a detailed derivation of the magnetic vector potential has previously not been published.

In this paper we wish to calculate the magnetic vector potential and subsequently the magnetic flux density at any point in a two dimensional space resulting from a Halbach cylinder.

Once the analytical solution for the magnetic flux density has been obtained we will proceed to calculate the force and torque between two concentric Halbach cylinders.

For $p = 1$ and a relative permeability of 1 the more complicated problem of computing the torque between two finite length concentric Halbach cylinders has been considered \cite{Mhiochain_1999}, and it is shown that a 
torque arises due to end effects. However, neither the field nor the torque is evaluated explicitly. Below we show that for special values of $p$ a nonzero force and torque may arise even in the two dimensional case.

\section{Defining the magnetostatic problem}
The problem of finding the magnetic vector potential and the magnetic flux density for a Halbach cylinder is defined in terms of the magnetic vector potential equation through the relation between the magnetic flux 
density, $\mathbf{B}$, and the magnetic vector potential, $\mathbf{A}$,
\begin{eqnarray}
\mathbf{B} = \mathbf{\nabla{}}\times{}\mathbf{A}~.\label{Eq.B_curl_A}
\end{eqnarray}

If there are no currents present it is possible to express the magnetic vector potential as
\begin{eqnarray}
-\mathbf{\nabla{}}^2\mathbf{A} = \mathbf{\nabla{}}\times{}\mathbf{B}_{\mathrm{rem}}~.
\end{eqnarray}

For the two dimensional case considered here the vector potential only has a $z$-component, $A_z$, and the above equation, using Eq. (\ref{Eq.Halbach_magnetization}), is reduced to
\begin{eqnarray} \label{Eq.Vector_equation}
-\mathbf{\nabla{}}^2 A_z(r,\phi) = \frac{ B_{\mathrm{rem}} }{r}\;(p + 1)\; \textrm{sin}(p \phi)\;.
\end{eqnarray}
This differential equation constitutes the magnetic vector potential problem and must be solved. In the air region of the problem the right hand side reduces to zero as here $B_{\mathrm{rem}} = 0$.

Once $A_z$ has been determined Eq. (\ref{Eq.B_curl_A}) can be used to find the magnetic flux density. Afterwards the magnetic field, $\mathbf{H}$, can be found through the relation
\begin{eqnarray}\label{Eq.Governing_equation}
\mathbf{B} = \mu{}_{0}\mu{}_{r}\mathbf{H}+ \mathbf{B}_{\mathrm{rem}}~,
\end{eqnarray}
where $\mu_r$ is the relative permeability assumed to be isotropic and independent of $\mathbf{B}$ and $\mathbf{H}$. This is generally the case for hard permanent magnetic materials.

\subsection{Geometry of the problem}
\begin{figure}[!t]
   \includegraphics[width=\columnwidth]{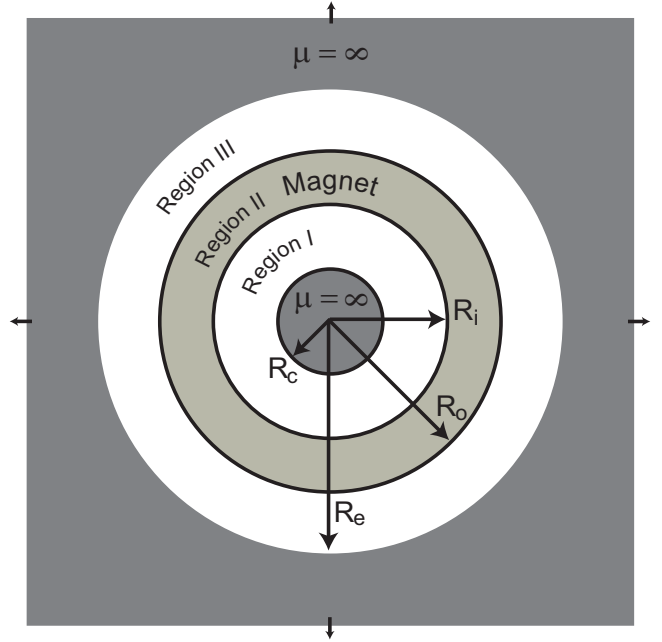}
      \caption{A Halbach cylinder with inner radius $\Rin$ and outer radius $\Rou$ enclosing an infinitely permeable cylinder with radius $\Rce$ while itself being enclosed by another infinitely permeable cylinder with
      inner radius $\Ren$ and infinite outer radius. The regions marked I and III are air gaps.}
       \label{Fig.Analytical_drawing}
\end{figure}

Having found the equation governing the magnetostatic problem of the Halbach cylinder we now take a closer look at the geometry of the problem. Following the approach of Xia et al. \cite{Xia_2004} we will start by solving 
the problem of a Halbach cylinder enclosing a cylinder of an infinitely permeable soft magnetic material, while at the same time itself being enclosed by another such cylinder. This is the situation depicted in Fig. 
\ref{Fig.Analytical_drawing}. This configuration is important for e.g. motor applications. The Halbach cylinder has an inner radius of $\Rin$ and an outer radius of $\Rou$ and the inner infinitely permeable cylinder has a 
radius of $\Rce$ while the outer enclosing cylinder has a inner radius of $\Ren$ and an infinite outer radius. Later in this paper we will solve the magnetostatic problem of the Halbach cylinder in air by letting $\Rce 
\rightarrow 0$ and $\Ren \rightarrow \infty$. The use of the soft magnetic cylinders results in a well defined set of boundary equations as will be shown later. Of course one can also solve directly for the Halbach 
cylinder in air using the boundary conditions specific for this case.

When solving the magnetostatic problem three different expressions for the magnetic vector potential, field and flux density will be obtained, one for each of the three different regions shown in Fig. 
\ref{Fig.Analytical_drawing}. The geometry of the problem results in six boundary conditions. The requirement is that the radial component of $\mathbf{B}$ and the parallel component of $\mathbf{H}$ are continuous across 
boundaries, i.e.
\begin{eqnarray}
H^{I}_{\phi} \;\;=& 0 &|\; r=\Rce \nonumber \\
B^{I}_{r} \;\;=& B^{II}_{r} &|\; r=\Rin \nonumber \\
H^{I}_{\phi} \;\;=& H^{II}_{\phi} &|\; r=\Rin \nonumber \\
B^{III}_{r} \;\;=& B^{II}_{r} &|\; r=\Rou \nonumber \\
H^{III}_{\phi} \;\;=& H^{II}_{\phi} &|\; r=\Rou \nonumber \\
H^{III}_{\phi} \;\;=& 0 &|\; r=\Ren \;.\label{Eq.Boundary_conditions}
\end{eqnarray}
The two equations for $H_{\phi}=0$ come from the fact that the soft magnetic material has an infinite permeability.

\subsection{Solution for the vector potential}
The solution to the vector potential equation, Eq. (\ref{Eq.Vector_equation}), is the sum of the solution to the homogenous equation and a particular solution. The solution is
\begin{equation} \label{Eq.Vector_solution_rough}
A_z (r,\phi) = \sum_{n=1}^{\infty}(\Acon{}_{n}r^n + \Bcon{}_{n}r^{-n})\textrm{sin}(n\phi) + B_{\mathrm{rem}} \frac{r}{p - 1}\textrm{sin}(p\phi)\;,
\end{equation}
where $\Acon{}_{n}$ and $\Bcon{}_{n}$ are constants that differ for each different region and that are different for each $n$. Using the boundary conditions for the geometry defined above one can show that these are only 
nonzero for $n=p$.

Thus the solution for the defined geometry becomes
\begin{equation} \label{Eq.Vector_solution}
A_z (r,\phi) = (\Acon{}r^p + \Bcon{}r^{-p})\textrm{sin}(p\phi) + B_{\mathrm{rem}} \frac{r}{p - 1}\textrm{sin}(p\phi)\;,
\end{equation}
where $\Acon{}$ and $\Bcon{}$ are constants that differ for each different region and that are determined by boundary conditions.

The solution is not valid for $p = 1$. For this special case the solution to Eq. (\ref{Eq.Vector_equation}) is instead
\begin{equation}\label{Eq.Vector_solution_p_1}
A_z (r,\phi) = (\Acon{}r + \Bcon{}r^{-1})\textrm{sin}(\phi) - B_{\mathrm{rem}} r\textrm{ln}(r)\textrm{sin}(\phi)\;,
\end{equation}
where $\Acon{}$ and $\Bcon{}$ are defined like for Eq. (\ref{Eq.Vector_solution}).

Note that for $p=0$ we have that $B_{\mathrm{rem},r} = B_{\mathrm{rem}}$ and $B_{\mathrm{rem},\phi} = 0$ in Eq. (\ref{Eq.Halbach_magnetization}). This means that $A_z = 0$ and consequently $\mathbf{B}$ is zero everywhere. 
The magnetic field, $\mathbf{H}$, however, will be nonzero inside the magnetic material itself, i.e. in region II, but will be zero everywhere else.

We now derive the constants in Eq. (\ref{Eq.Vector_solution}) and (\ref{Eq.Vector_solution_p_1}) directly from the boundary conditions.

\section{Deriving the vector potential constants}
The constants of the vector potential equation can be derived from the boundary conditions specified in Eq. (\ref{Eq.Boundary_conditions}). We first derive the constants for the case of $p \neq 1$.

First we note that the magnetic flux density and the magnetic field can be calculated from the magnetic vector potential
\begin{eqnarray}\label{Eq.Magnetic_flux_definition_derivatives}
B_r &=&\frac{1}{r}\frac{\pd A_z}{\pd \phi}\nonumber\\
B_\phi &=& -\frac{\pd A_z}{\pd r}\nonumber\\
H_r    &=& \frac{1}{\mu_0 \mu_r}(B_r - B_{\mathrm{rem},r})\nonumber\\
H_\phi &=& \frac{1}{\mu_0 \mu_r}(B_\phi - B_{\mathrm{rem},\phi})\;.
\end{eqnarray}
Performing the differentiation gives
\begin{eqnarray}
B_r    &=& \left[ p \Acon{}r^{p-1} + p \Bcon{}r^{-p-1} + B_{\mathrm{rem}} \frac{p}{p - 1}\right]\textrm{cos}(p\phi) \nonumber\\
B_\phi &=& \left[-p \Acon{}r^{p-1} + p \Bcon{}r^{-p-1} - B_{\mathrm{rem}} \frac{1}{p - 1}\right]\textrm{sin}(p\phi) \nonumber\\
H_r    &=& \left[\frac{p}{\mu_0 \mu_r}( \Acon{}r^{p-1} + \Bcon{}r^{-p-1}) \right.\nonumber\\
        &&\left. + \frac{B_{\mathrm{rem}}}{\mu_r \mu_0}\left(\frac{p}{p - 1}-1\right)\right]\textrm{cos}(p\phi) \nonumber\\
H_\phi &=& \left[\frac{p}{\mu_0 \mu_r}(-\Acon{}r^{p-1} + \Bcon{}r^{-p-1}) \right.\nonumber\\
        &&\left.- \frac{B_{\mathrm{rem}}}{\mu_r \mu_0}\left(\frac{1}{p - 1} - 1\right)\right]\textrm{sin}(p\phi)\;.\label{Eq.Magnetic_flux_definition}
\end{eqnarray}

Using the radial component of the magnetic flux density and the tangential component of the magnetic field in the set of boundary equations we get a set of six equations containing the six unknown constants, two for each 
region. The constants $\Acon{}$ and $\Bcon{}$ will be termed $\Acon{I}$ and $\Bcon{I}$ in region I, $\Acon{II}$ and $\Bcon{II}$ in region II, and $\Acon{III}$ and $\Bcon{III}$ in region III.

Introducing the following new constants
\begin{eqnarray}
a &=& \frac{\Ren^{2p} - \Rou^{2p}}{\Ren^{2p} + \Rou^{2p}}\nonumber\\
b &=& -\frac{\Rin^{2p} - \Rce^{2p}}{\Rin^{2p} + \Rce^{2p}}\;,\label{Eq.a_b_constants}
\end{eqnarray}
the constants are determined to be
\begin{eqnarray}
\Bcon{II} = -\frac{\Rou^{1-p} - \Rin^{1-p}}{\frac{\mu_r a -1}{\mu_r a +1}\Rou^{-2p} - \frac{\mu_r b-1}{\mu_r b +1}\Rin^{-2p}}\frac{\Brem}{p-1}\;,
\end{eqnarray}

and
\begin{eqnarray}
\Acon{I} &=& \frac{\Bcon{II}}{\Rin^{2p} + \Rce^{2p}}\left(1-\frac{\mu_r b-1}{\mu_r b +1}\right)\nonumber\\
\Bcon{I} &=& \Acon{I} \Rce^{2p}\nonumber\\
\Acon{II} &=& -\Bcon{II}\frac{\mu_r a -1}{\mu_r a +1}\Rou^{-2p} - \frac{\Brem}{p-1}\Rou^{1-p}\nonumber\\
\Acon{III} &=& \frac{\Bcon{II}}{\Rou^{2p} + \Ren^{2p}}\left(1-\frac{\mu_r a-1}{\mu_r a+1}\right)\nonumber\\
\Bcon{III} &=& \Acon{III} \Ren^{2p}\;.\label{Eq.C_iron}
\end{eqnarray}
Using these constants in Eq. (\ref{Eq.Vector_solution}) and (\ref{Eq.Magnetic_flux_definition}) allows one to calculate the magnetic vector potential, the magnetic flux density and the magnetic field respectively.

The constants are not valid for $p = 1$. The solution for this case will be derived in a later section.

\subsection{Halbach cylinder in air}
We can find the solution for a Halbach cylinder in air if we look at the solution for $\Ren \rightarrow \infty$ and $\Rce \rightarrow 0$. Looking at the previous expression for the constants $a$ and $b$ we see that
\begin{eqnarray}
\textrm{for } p > 1 &:&
\begin{array}{l}
a \rightarrow 1\\
b \rightarrow -1\\
\end{array}
\nonumber\\
\textrm{for } p < 0 &:&
\begin{array}{l}
a \rightarrow -1\\
b \rightarrow 1\\
\end{array}
\end{eqnarray}
in the limit defined above.

This means that the constant $\Bcon{II}$ now becomes
\begin{eqnarray}
\Bcon{II} = \left\{
\begin{array}{l}
-\frac{\Rou^{1-p} - \Rin^{1-p}}{\frac{\mu_r-1}{\mu_r+1}\Rou^{-2p} - \frac{\mu_r+1}{\mu_r-1}\Rin^{-2p}}\frac{\Brem}{p-1} \qquad p > 1\\
-\frac{\Rou^{1-p} - \Rin^{1-p}}{\frac{\mu_r+1}{\mu_r-1}\Rou^{-2p} - \frac{\mu_r-1}{\mu_r+1}\Rin^{-2p}}\frac{\Brem}{p-1} \qquad p < 0
\end{array}
\right.
\end{eqnarray}

and the remaining constants for $p>1$ become
\begin{align}
\left[\begin{array}{l}
\Acon{I} \\
\Bcon{I} \\
\Acon{II} \\
\Acon{III} \\
\Bcon{III} \\
\end{array}\right] &=&
\left\{
\begin{array}{l}
\Bcon{II}\Rin^{-2p}\left(1-\frac{\mu_r+1}{\mu_r-1}\right)\\
0\\
-\Bcon{II}\frac{\mu_r-1}{\mu_r+1}\Rou^{-2p} - \frac{\Brem}{p-1}\Rou^{1-p}\\
0\\
\Bcon{II}\left(1-\frac{\mu_r-1}{\mu_r+1}\right)\\
\end{array}
\right. \label{Eq.C_air_p_gt_0}
\end{align}
while for $p<0$ they become
\begin{align}
\left[\begin{array}{l}
\Acon{I} \\
\Bcon{I} \\
\Acon{II} \\
\Acon{III} \\
\Bcon{III} \\
\end{array}\right] &=&
\left\{
\begin{array}{l}
0\\
\Bcon{II}\left(1-\frac{\mu_r-1}{\mu_r+1}\right) \\
-\Bcon{II}\frac{\mu_r+1}{\mu_r-1}\Rou^{-2p} - \frac{\Brem}{p-1}\Rou^{1-p}\\
\Bcon{II}\Rou^{-2p}\left(1-\frac{\mu_r+1}{\mu_r-1}\right) \\
0 \\
\end{array}
\right. \label{Eq.C_air_mur_all}
\end{align}

This is the solution for a Halbach cylinder in air. Note that the solution is only valid for $\mu_r \neq 1$. In the special case of $\mu_r = 1$ the constants can be reduced even further.

\subsection{Halbach cylinder in air and $\mu_r = 1$}
We now look at the special case of a Halbach cylinder in air with $\mu_r = 1$. This is a relevant case as e.g. the highest energy density type of permanent magnet produced today, the so-called neodymium-iron-boron (NdFeB) 
magnets, have a relative permeability very close to one: $\mu_r = 1.05$ \cite{Standard}.

Using the approximation of $\mu_r \rightarrow 1$ for a Halbach cylinder in air reduces the constant $\Bcon{II}$ to
\begin{eqnarray}
\Bcon{II} = 0 ~.
\end{eqnarray}

The remaining constants depend on whether the Halbach cylinder produces an internal or external field.

For the internal field case, $p > 1$, the constant $\Acon{II}$ will be given by
\begin{eqnarray}
\Acon{II} = -\frac{\Brem}{p-1}\Rou^{1-p} ~.
\end{eqnarray}

The constant $\Acon{I}$ determining the field in the inner air region is equal to
\begin{eqnarray}
\Acon{I} = \frac{\Brem}{p - 1}\left(\Rin^{1-p} - \Rou^{1-p}\right)\;.
\end{eqnarray}
The remaining constants, $\Bcon{I}$, $\Acon{III}$ and $\Bcon{III}$ are zero.

Using Eq. (\ref{Eq.Magnetic_flux_definition}) the two components of the magnetic flux density in both the cylinder bore, region I, and in the magnet, region II, can be found.
\begin{eqnarray}
B^{I}_{r}    &=& \frac{\Brem p}{p - 1}\left(1 - \left(\frac{\Rin}{\Rou}\right)^{p-1} \right)\times\nonumber\\&&\left(\frac{r}{\Rin}\right)^{p-1}\textrm{cos}(p\phi)\nonumber\\
B^{I}_{\phi} &=& -\frac{\Brem p}{p - 1}\left(1-\left(\frac{\Rin}{\Rou}\right)^{p-1}\right)\times\nonumber\\&&\left(\frac{r}{\Rin}\right)^{p-1}\textrm{sin}(p\phi)\nonumber\\
B^{II}_{r}   &=&  \frac{\Brem p}{p - 1}\left(1 - \left(\frac{r}{\Rou}\right)^{p-1}\right)\textrm{cos}(p\phi)\nonumber\\
B^{II}_{\phi}&=& -\frac{\Brem}{p - 1}  \left(1- p\left(\frac{r}{\Rou}\right)^{p-1}\right)\textrm{sin}(p\phi)\;.\label{Eq.B_internal_air}
\end{eqnarray}

Considering now the external field case, $p < 0$, the constant $\Acon{II}$ is given by
\begin{eqnarray}
\Acon{II} = -\frac{\Brem}{p-1}\Rin^{1-p}\;.
\end{eqnarray}

The constant $\Acon{III}$ determining the field in the outer air region is given by
\begin{eqnarray}
\Acon{III} = \frac{\Brem}{p - 1}\left(\Rou^{p-1} - \Rin^{p-1}\right)\;.
\end{eqnarray}
The remaining constants, $\Acon{I}$, $\Bcon{I}$ and $\Bcon{III}$ are zero.

Again using Eq. (\ref{Eq.Magnetic_flux_definition}) we find the two components of the magnetic flux density in region II and III to be
\begin{eqnarray}
B^{III}_{r}    &=&  \frac{\Brem p}{p - 1}\left(1 - \left(\frac{\Rin}{\Rou}\right)^{-p+1} \right)\times\nonumber\\&&\left(\frac{\Rou}{r}\right)^{-p+1}\textrm{cos}(p\phi)\nonumber\\
B^{III}_{\phi} &=& -\frac{\Brem p}{p - 1}\left(1 - \left(\frac{\Rin}{\Rou}\right)^{-p+1} \right)\times\nonumber\\&&\left(\frac{\Rou}{r}\right)^{-p+1}\textrm{sin}(p\phi)\nonumber\\
B^{II}_{r}     &=& \frac{\Brem p}{p - 1}\left(1 - \left(\frac{\Rin}{r}\right)^{-p+1}\right)\textrm{cos}(p\phi)\nonumber\\
B^{II}_{\phi}  &=& -\frac{\Brem}{p - 1}  \left(1- p\left(\frac{\Rin}{r}\right)^{-p+1}\right)\textrm{sin}(p\phi)\;.\label{Eq.B_external_air}
\end{eqnarray}
The equations for $B^{III}_{r}$ and $B^{III}_{\phi}$ are identical to the expressions for $B^{I}_{r}$ and $B^{I}_{\phi}$ in Eq. (\ref{Eq.B_internal_air}) except for a minus sign in both equations.


\subsection{The constants for a $p = 1$ Halbach cylinder}
Having determined the solution to the vector potential equation and found the constants in the expression for the magnetic flux density and the magnetic vector potential for a Halbach cylinder both in air and enclosed by a 
soft magnetic cylinder for all cases except $p=1$  we now turn to this specific case. This case is shown in Fig \ref{Fig.Magnetization_drawing}. We have already shown that the solution to the vector potential problem for 
this case is given by Eq. (\ref{Eq.Vector_solution_p_1}). The boundary conditions are the same as previous, i.e. they are given by Eq. (\ref{Eq.Boundary_conditions}).

In order to find the constants the components of the magnetic field and the magnetic flux density must be calculated for $p=1$ as the boundary conditions relate to these fields. Using Eq. 
(\ref{Eq.Magnetic_flux_definition_derivatives}) we obtain
\begin{eqnarray}
B_r    &=& [\Acon{} + \Bcon{}r^{-2} - \Brem\textrm{ln}(r)]\textrm{cos}(\phi)\nonumber\\
B_\phi &=& [-\Acon{} + \Bcon{} r^{-2} + \Brem(\textrm{ln}(r)+1)]\textrm{sin}(\phi)\nonumber\\
H_r    &=& \frac{1}{\mu_0 \mu_r}\left[\Acon{} + \Bcon{}r^{-2} - \Brem(\textrm{ln}(r)+1)\right]\textrm{cos}(\phi)\nonumber\\
H_\phi &=& \frac{1}{\mu_0 \mu_r}\left[-\Acon{} + \Bcon{} r^{-2} + \Brem\textrm{ln}(r)\right]\textrm{sin}(\phi)\;.\label{Eq.B_and_H_for_p_1}
\end{eqnarray}

Using these expressions for the magnetic flux density and the magnetic field we can again write a set of six equations through which we can determine the six constants, two for each region.

Reintroducing the two constants from Eq. (\ref{Eq.a_b_constants})
\begin{eqnarray}
a &=& \frac{\Ren^2-\Rou^2}{\Ren^2+\Rou^2}\nonumber\\
b &=& -\frac{\Rin^2-\Rce^2}{\Rin^2+\Rce^2}~,
\end{eqnarray}
the following equations for the constants are obtained:
\begin{eqnarray}
\Acon{I} &=&  \frac{\Bcon{II}}{\Rin^{2} + \Rce^{2}}\left(1-\frac{\mu_r b -1}{\mu_r b+1}\right)\nonumber\\
\Bcon{I} &=& \Acon{I} \Rce^{2}\nonumber\\
\Acon{II} &=& -\Bcon{II}\frac{\mu_r a-1}{\mu_r a+1} \Rou^{-2} + \Brem\;\textrm{ln}(\Rou)\nonumber\\
\Bcon{II} &=& -\left(\frac{a\mu_r-1}{a\mu_r+1} \Rou^{-2} - \frac{\mu_r b-1}{\mu_r b+1} \Rin^{-2}\right)^{-1}\times\nonumber\\&& \Brem\;\textrm{ln}\left(\frac{\Rin}{\Rou}\right)\nonumber\\
\Acon{III} &=& \frac{\Bcon{II}}{\Ren^2+\Rou^2} \left(1-\frac{\mu_r a-1}{\mu_r a+1}\right)\nonumber\\
\Bcon{III} &=& \Acon{III} \Ren^{2}\;.
\end{eqnarray}
We see that the constants $\Acon{I}$, $\Bcon{I}$, $\Acon{III}$ and $\Bcon{III}$ are identical to the constants in Eq. (\ref{Eq.C_iron}).

The magnetic flux density and the magnetic field can now be found through the use of Eq. (\ref{Eq.B_and_H_for_p_1}).

\subsection{Halbach cylinder in air, $p = 1$}
We can find the solution for a $p=1$ Halbach cylinder in air if we look at the solution for $\Ren \rightarrow \infty$ and $\Rce \rightarrow 0$. In this limit the previously introduced constants are reduced to
\begin{eqnarray}
a &\rightarrow& 1\nonumber\\
b &\rightarrow& -1\;.
\end{eqnarray}

The expressions for the constants can then be reduced to
\begin{eqnarray}
\Acon{I} &=& \Bcon{II}\Rin^{-2}\left(1-\frac{\mu_r+1}{\mu_r-1}\right)\nonumber\\
\Bcon{I} &=& 0\nonumber\\
\Acon{II} &=& -\Bcon{II}\frac{\mu_r-1}{\mu_r+1} \Rou^{-2} + \Brem\;\textrm{ln}(\Rou)\nonumber\\
\Bcon{II} &=& -\left(\frac{\mu_r-1}{\mu_r+1} \Rou^{-2} - \frac{\mu_r+1}{\mu_r-1} \Rin^{-2}\right)^{-1}\times\nonumber\\&& \Brem\;\textrm{ln}\left(\frac{\Rin}{\Rou}\right)\nonumber\\
\Acon{III} &=& 0\nonumber\\
\Bcon{III} &=& \Bcon{II}\left(1-\frac{\mu_r-1}{\mu_r+1}\right)\;.
\end{eqnarray}
Again we see that the constants $\Acon{I}$, $\Bcon{I}$, $\Acon{III}$ and $\Bcon{III}$ are equal to the constants in Eq. (\ref{Eq.C_air_p_gt_0}).
This solution is valid for all $\mu_r$ except $\mu_r = 1$.

Combining the above constants with Eq. (\ref{Eq.B_and_H_for_p_1}) we see that the magnetic flux density in the cylinder bore is a constant, and that its magnitude is given by
\begin{align}
||\mathbf{B}^I|| &=& \left(\frac{\mu_r-1}{\mu_r+1} \Rou^{-2} - \frac{\mu_r+1}{\mu_r-1} \Rin^{-2}\right)^{-1} \nonumber\\
                  &&\times\left(\frac{\mu_r+1}{\mu_r-1}-1\right)\Rin^{-2} \Brem\;\textrm{ln}\left(\frac{\Rin}{\Rou}\right)~,
\end{align}
for $\mu_r \neq 1$.

\subsection{Halbach cylinder in air, $p=1$ and $\mu_r = 1$}
For the special case of $\mu_r = 1$ for a $p=1$ Halbach cylinder in air the constants can be reduced further to
\begin{eqnarray}
\Acon{I} &=& \Brem\;\textrm{ln}\left(\frac{\Rou}{\Rin}\right)\nonumber\\
\Acon{II} &=& \Brem\;\textrm{ln}(\Rou)\nonumber\\
\Bcon{I},\Bcon{II},\Acon{III},\Bcon{III} &=& 0\;.
\end{eqnarray}

Combining the above constants with Eq. (\ref{Eq.B_and_H_for_p_1}) one can find the magnetic flux density in the bore, region I, and in the magnet, region II,
\begin{eqnarray}
B^{I}_{r} &=& \Brem\;\textrm{ln}\left(\frac{\Rou}{\Rin}\right)\textrm{cos}(\phi)\nonumber\\
B^{I}_{\phi} &=& -\Brem\;\textrm{ln}\left(\frac{\Rou}{\Rin}\right)\textrm{sin}(\phi)\nonumber\\
B^{II}_{r} &=& \Brem\textrm{ln}\left(\frac{\Rou}{r}\right)\textrm{cos}(\phi)\nonumber\\
B^{II}_{\phi} &=& -\Brem\left(\textrm{ln}\left(\frac{\Rou}{r}\right) -1\right)\textrm{sin}(\phi)\;.\label{Eq.B_internal_air_p_1}
\end{eqnarray}
As for the case of $\mu_r \neq 1$ the magnetic flux density in the cylinder bore is a constant. The magnitude of the magnetic flux density in the bore is given by
\begin{eqnarray}
||\mathbf{B}^I|| &=& \Brem\;\textrm{ln}\left(\frac{\Rou}{\Rin}\right)\;.
\end{eqnarray}
which we recognize as the well known Halbach formula \cite{Halbach_1980}.


\subsection{Validity of the solutions}
To show the validity of the analytical solutions we compare these with a numerical calculation of the vector potential and the magnetic flux density.

\begin{figure}
\subfigure{\includegraphics[width=\columnwidth]{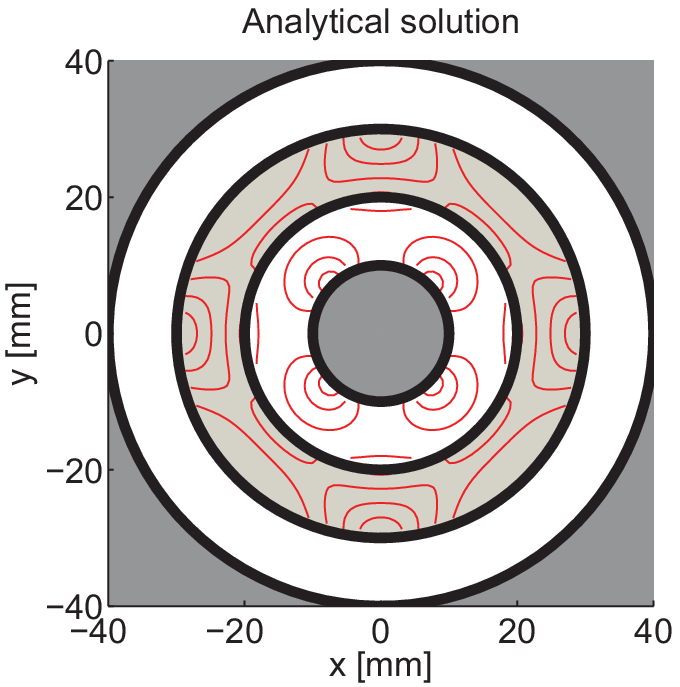}}
\subfigure{\includegraphics[width=\columnwidth]{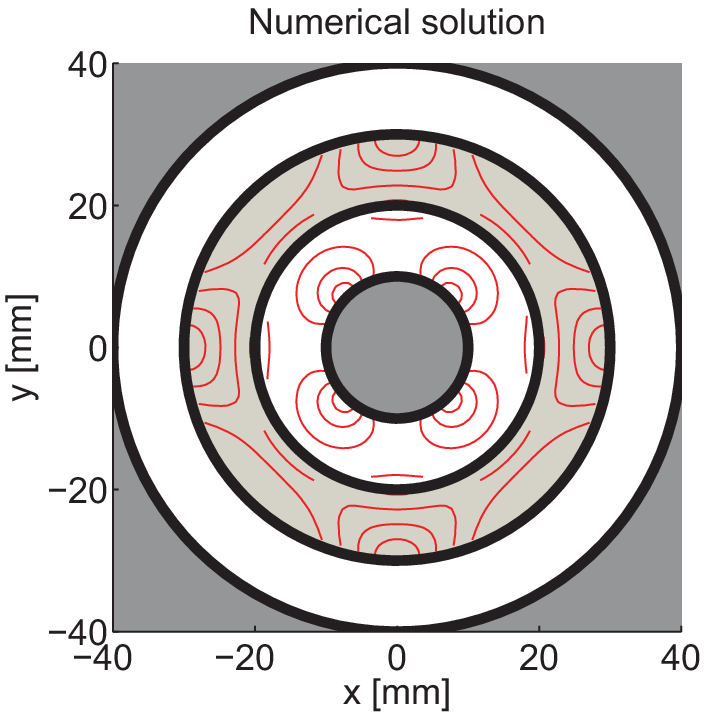}}
\caption{(Color online) Comparing the analytical solution as given by Eq. (\ref{Eq.Magnetic_flux_definition}) and (\ref{Eq.C_iron}) with a numerical solution computed using Comsol. Shown are contours of $||\mathbf{B}|| =
[0.3,0.5,0.7,0.9]$ T for an internal field $p = 2$ enclosed Halbach cylinder with dimensions $\Rce = 10$ mm, $\Rin = 20$ mm, $\Rou = 30$ mm, $\Ren = 40$ mm, and $\Brem = 1.4$ T, $\mu_r  = 1.05$. The solutions are seen to
be identical. The shaded areas in the figures correspond to the similar shaded areas in Fig. \ref{Fig.Analytical_drawing}.}\label{Fig.Compare_Bx}
\end{figure}

\begin{figure}
\subfigure{\includegraphics[width=\columnwidth]{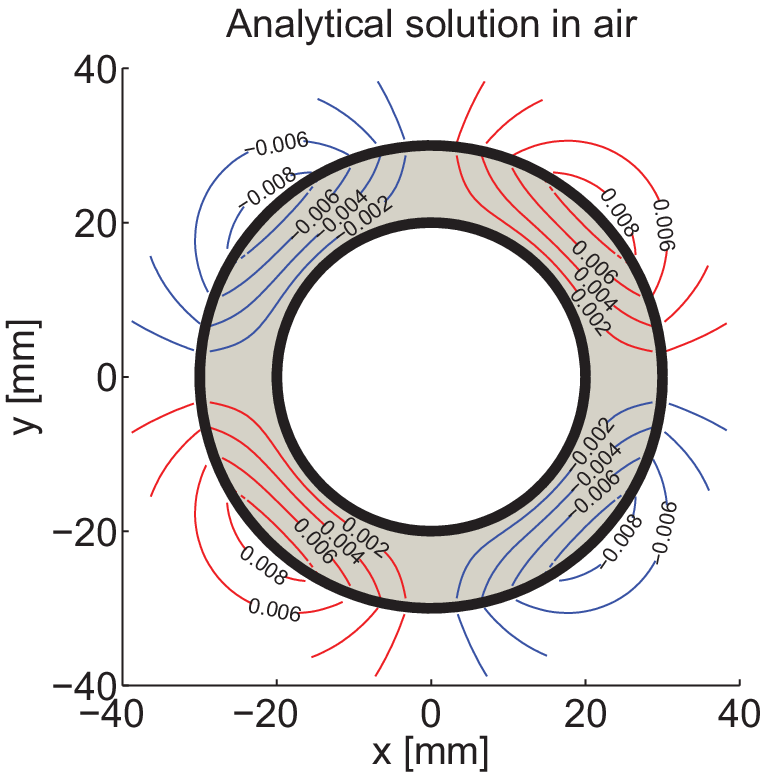}}
\subfigure{\includegraphics[width=\columnwidth]{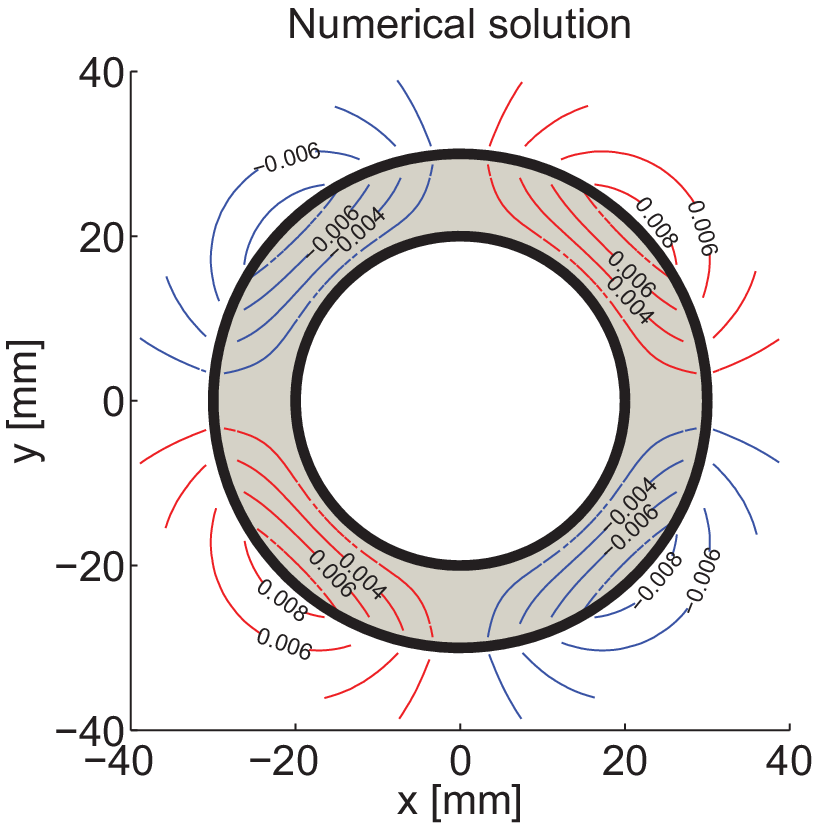}}
\caption{(Color online) Comparing the analytical solution as given by Eqs. (\ref{Eq.Vector_solution}) and (\ref{Eq.C_air_mur_all}) with a numerical solution computed using Comsol. Shown are contours of $A_z = \pm[0.002,
0.004, 0.006, 0.008]$V s m$^{-1}$ for an external field $p = -2$ Halbach cylinder in air with dimensions $\Rin = 20$ mm, $\Rou = 30$ mm and $\Brem = 1.4$ T, $\mu_r  = 1.05$. The red contours are positive values of $A_z$
while the blue are negative values. As with Fig. \ref{Fig.Compare_Bx} the solutions are seen to be identical.}\label{Fig.Compare_Az}
\end{figure}

We have chosen to show a comparison between the expressions derived in this paper and numerical calculations for two selected cases. These are shown in Fig. \ref{Fig.Compare_Bx} and \ref{Fig.Compare_Az}.

In Fig. \ref{Fig.Compare_Bx} the magnitude of the magnetic flux density is shown for a enclosed Halbach cylinder. Also shown in Fig. \ref{Fig.Compare_Bx} is a numerical calculation done using the commercially available 
finite element multiphysics program, \emph{Comsol Multiphysics} \cite{Comsol}. The Comsol Multiphysics code has previously been validated through a number of NAFEMS (National Agency for Finite Element Methods and 
Standards) benchmark studies \cite{Comsol_2005}. As can be seen the analytical solution closely matches the numerical solution.

In Fig. \ref{Fig.Compare_Az} we show the magnetic vector potential, $A_z$, as calculated using Eqs. (\ref{Eq.Vector_solution}) and (\ref{Eq.C_air_mur_all}) compared with a numerical Comsol simulation. As can be seen the 
analytical solution again closely matches the numerical solution.

We have also tested the expressions for the magnetic flux density given by Xia \emph{et. al.} (2004) \cite{Xia_2004} and compared them with those derived in this paper and with numerical calculations. Unfortunately the 
equations given by Xia \emph{et. al.} (2004) \cite{Xia_2004} contain erroneous expressions for the magnetic flux density of a Halbach cylinder in air with $\mu_r = 1$ as well as for the expression for a Halbach cylinder 
with internal field enclosed by soft magnetic material.

\section{Force between two concentric Halbach cylinders}
\begin{figure}[!t]
\includegraphics[width=\columnwidth]{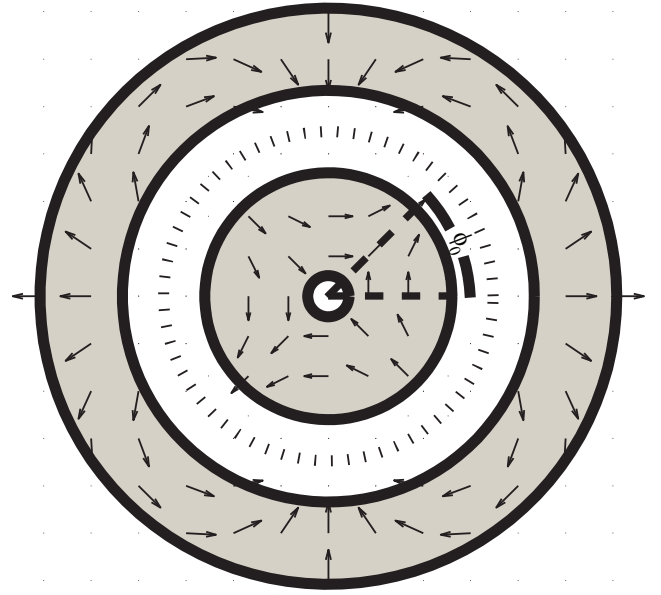}
\caption{An example of a concentric Halbach cylinder configuration for which the force and torque is calculated. The outer magnet has $p=2$ while the inner magnet is a $p=-2$. The inner magnet has also been rotated an
angle of $\phi_0 = 45^{\circ}$. The dotted circle indicates a possible integration path.}\label{Fig.Magnetization_Two_Halbachs}
\end{figure}

Having found the expressions for the magnetic vector potential and the magnetic flux density for a Halbach cylinder we now turn to the problem of calculating the force between two concentric Halbach cylinders, e.g. a 
situation as shown in Fig. \ref{Fig.Magnetization_Two_Halbachs}. In a later section we will calculate the torque for the same configuration. This configuration is interesting for e.g. motor applications and drives as well 
as applications where the magnetic flux density must be turned ``on'' and ``off'' without the magnet being displaced in space \cite{Tura_2007}.

The force between the two Halbach cylinders can be calculated by using the Maxwell stress tensor, $\overleftrightarrow{\mathbf{T}}$, formulation. The force per unit length is given by
\begin{eqnarray}
\mathbf{F} &=& \frac{1}{\mu_0} \oint_S\overleftrightarrow{\mathbf{T}}\cdot dS \;. \label{Eq.Maxwell_Force_Def} 
\end{eqnarray}

The Cartesian components of the force are given by
\begin{eqnarray}
F_x &=& \frac{1}{\mu_0} \oint_S(T_{xx}n_x+T_{xy}n_y)ds\nonumber\\
F_y &=& \frac{1}{\mu_0} \oint_S(T_{yy}n_y+T_{yx}n_x)ds\;,
\end{eqnarray}
where $n_x$ and $n_y$ are the Cartesian components of the outwards normal to the integration surface and where $T_{xx}$, $T_{yy}$ and $T_{xy}$ are the components of the Maxwell stress tensor which are given by
\begin{eqnarray}
T_{xx} &=& B_x^2-\frac{1}{2}(B_x^2+B_y^2)\nonumber\\
T_{yy} &=& B_y^2-\frac{1}{2}(B_x^2+B_y^2)\nonumber\\
T_{xy},\; T_{yx} &=& B_xB_y\;.
\end{eqnarray}

When using the above formulation to calculate the force a closed integration surface in free space that surrounds the object must be chosen. As this is a two dimensional problem the surface integral is reduced to a line 
integral along the air gap between the magnets. If a circle of radius $r$ is taken as the integration path, the Cartesian components of the outwards normal are given by
\begin{eqnarray}
n_x &=& \textrm{cos}(\phi)\nonumber\\
n_y &=& \textrm{sin}(\phi)\;.
\end{eqnarray}
Expressing the Cartesian components through the polar components as
\begin{eqnarray}
B_x &=& B_r\textrm{cos}(\phi) - B_\phi\textrm{sin}(\phi)\nonumber \\
B_y &=& B_r\textrm{sin}(\phi) + B_\phi\textrm{cos}(\phi)\;,\label{Eq.Bx_By_from_Br_Bt}
\end{eqnarray}
the relation for computing the force per unit length becomes
\begin{align}\label{Eq.Force_integral}
F_x  &=& \frac{r}{\mu_0} \int_0^{2\pi}\left(\frac{1}{2}(B_r^2-B_\phi^2)\textrm{cos}(\phi) - B_r B_\phi \textrm{sin}(\phi)\right)\; d\phi\nonumber\\
F_y  &=& \frac{r}{\mu_0} \int_0^{2\pi}\left(\frac{1}{2}(B_r^2-B_\phi^2)\textrm{sin}(\phi) + B_r B_\phi \textrm{cos}(\phi)\right)\; d\phi\;,
\end{align}
where $r$ is some radius in the air gap. The computed force will turn out to be independent of the radius $r$ as expected.

We consider the scenario where the outer magnet is kept fixed and the internal magnet is rotated by an angle $\phi_0$, as shown in Fig. \ref{Fig.Magnetization_Two_Halbachs}. Both cylinders are centered on the same axis. 
Both of the cylinders are considered to be in air and have a relative permeability of one, $\mu_r = 1$, so that their magnetic flux density is given by Eqs. (\ref{Eq.B_internal_air}) and (\ref{Eq.B_external_air}) for $p 
\neq 1$. For $p=1$ Eq. (\ref{Eq.B_internal_air_p_1}) applies instead.

As $\mu_r = 1$ the magnetic flux density in the air gap between the magnets will be a sum of two terms, namely a term from the outer magnet and a term from the inner magnet. If the relative permeability were different from 
one the magnetic flux density of one of the magnets would influence the magnetic flux density of the other, and we would have to solve the vector potential equation for both magnets at the same time in order to find the 
magnetic flux density in the air gap.

Assuming the above requirements the flux density in the air gap is thus given by
\begin{eqnarray}
B_r    &=& B^{III}_{r,1}+B^{I}_{r,2}\nonumber\\
B_\phi &=& B^{III}_{\phi,1}+B^{I}_{\phi,2}\;,
\end{eqnarray}
where the second subscript refers to either of the two magnets. The inner magnet is termed ``1'' and the outer magnet termed ``2'', e.g. $R_{\textrm{o},1}$ is the inner magnets outer radius. The integer $p_1$ thus refers 
to the inner magnet and $p_2$ to the outer magnet.

There can only be a force between the cylinders if the inner cylinder produces an external field and the outer cylinder produces an internal field. Otherwise the flux density in the gap between the magnets will be produced 
solely by one of the magnets and the force will be zero.

Performing the integrals in Eq. (\ref{Eq.Force_integral}) one only obtains a nonzero solution for $p_1 = 1-p_2$ and $p_2>1$. In this case the solution is
\begin{eqnarray}\label{Eq.Force_solution}
F_x  &=& \frac{2\pi}{\mu_0}K\textrm{cos}(p_1\phi_0)\nonumber\\
F_y  &=& \frac{2\pi}{\mu_0}K\textrm{sin}(p_1\phi_0)\;,
\end{eqnarray}
where $K$ is a constant given by
\begin{align}
K &=& \Brem{}_{,1}\Brem{}_{,2}(R_{\textrm{i},2}^{p_1}-R_{\textrm{o},2}^{p_1})(R_{\textrm{o},1}^{p_2}-R_{\textrm{i},1}^{p_2})\;.
\end{align}
Notice that the force is independent of $r$, as expected.

\begin{figure}[!t]
   \includegraphics[width=\columnwidth]{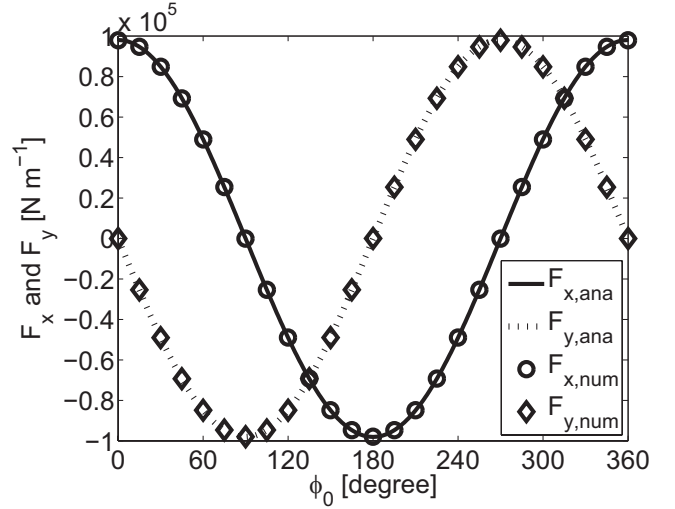}
      \caption{The two cartesian components of the force per unit length given by Eq. (\ref{Eq.Force_solution}) compared with a Comsol calculation for a system where the outer magnet has $p_2=2$, $R_{\textrm{i},2} = 45$
      mm, $R_{\textrm{o},2} = 75$ mm and $\Brem{}_{,2} = 1.4$ T and the inner magnet has $p_1=-1$, $R_{\textrm{i},1} = 15$ mm, $R_{\textrm{o},1} = 35$ mm and $\Brem{}_{,1} = 1.4$ T. The analytical expression is in
      excellent agreement with the numerical data. The force is per unit length as we consider a two dimensional system.}
       \label{Fig.Force_compare_pi_1_po_2}
\end{figure}

In Fig. \ref{Fig.Force_compare_pi_1_po_2} we compare the above equation with a numerical calculation of the force. The results are seen to be in excellent agreement. Notice that the forces never balance the magnets, i.e. 
when $F_{\mathrm{x}}$ is zero, $F_{\mathrm{y}}$ is nonzero and vice versa.

If $p_2 = 1$ the magnetic flux density produced by the outer magnet is not given by Eq. (\ref{Eq.B_internal_air}) but is instead given by Eq. (\ref{Eq.B_internal_air_p_1}). However this equation has the same angular 
dependence as Eq. (\ref{Eq.B_internal_air}) and thus the force will also be zero for this case.

\section{Torque between two concentric nested Halbach cylinders}
Having calculated the force between two concentric Halbach cylinders we now focus on calculating the torque for the same system.

The torque can also be calculated by using the Maxwell stress tensor, $\overleftrightarrow{\mathbf{T}}$, formulation. The torque per unit length is given by
\begin{eqnarray}
\mathbf{\tau} &=& \frac{1}{\mu_0} \oint_S \mathbf{r}\times\overleftrightarrow{\mathbf{T}}\cdot dS \nonumber\\
              &=& \frac{1}{\mu_0} \oint_S \mathbf{r}\left((\mathbf{B}\cdot\mathbf{n})\mathbf{B} - \frac{1}{2}\mathbf{B}^2\mathbf{n}\right)dS\;,
\end{eqnarray}
where again the integration surface is a closed loop in free space that surrounds the object. Again choosing a circle of radius $r$ as the integration path, the relation for computing the torque per unit length around the 
central axis becomes
\begin{eqnarray}
\mathbf{\tau} &=& \frac{1}{\mu_0}\int_0^{2\pi}r^2B_rB_\phi d\phi ~, \end{eqnarray}
where $B_r$ and $B_\phi$ are the radial and tangential components of the magnetic flux density in the air gap and $r$ is some radius in the air gap. Again the computed torque will be shown to be independent of the radius 
$r$ when $r$ varies between the inner and outer radii of the air gap.

We consider the same case as with the force calculation, i.e. the outer magnet is kept fixed, both magnets have the same axis, the internal magnet is rotated by an angle $\phi_0$ and both of the cylinders are considered to 
be in air and have a relative permeability of one. Again there can only be a torque between the cylinders if the inner cylinder produces an external field and the outer cylinder produces an internal field.

To find the torque per unit length we must thus integrate
\begin{align}\label{Eq.Torque_definition}
\mathbf{\tau} &=& \frac{1}{\mu_0}\int_0^{2\pi}r^2(B^{III}_{r,1}+B^{I}_{r,2})(B^{III}_{\phi,1}+B^{I}_{\phi,2}) d\phi ~.
\end{align}

This integration will be zero except when $p_1 = -p_2$. For this special case the integral gives
\begin{eqnarray}\label{Eq.Torque_solution}
\tau &=& \frac{2\pi}{\mu_0}\frac{ p_2^2}{1-p_2^2} K_1 K_2  \textrm{sin}(p_2\phi_0)\;,
\end{eqnarray}
where the constants $K_1$ and $K_2$ are given by
\begin{eqnarray}\label{Eq.Constants_torque_p_gt_1}
K_1 &=& \Brem{}_{,2}\left(R_{\textrm{i},2}^{1-p_2} - R_{\textrm{o},2}^{1-p_2}\right)\nonumber\\
K_2 &=& \Brem{}_{,1}\left(R_{\textrm{o},1}^{p_2+1} - R_{\textrm{i},1}^{p_2+1}\right)\;.
\end{eqnarray}

The validity of this expression will be shown in the next section. It is seen that there are $p_2$ periods per rotation.

For $p_2=1$ the expression for the magnetic flux density produced by the outer magnet is not given by Eq. (\ref{Eq.B_internal_air}) but instead by Eq. (\ref{Eq.B_internal_air_p_1}), and so we must look at this special case 
separately.

\subsection{The special case of $p_2 = 1$}
For the special case of a $p_2 = 1$ outer magnet the flux density produced by this magnet in the air gap will be given by Eq. (\ref{Eq.B_internal_air_p_1}). The external field produced by the inner magnet is still given by 
Eq. (\ref{Eq.B_external_air}).

Performing the integration defined in Eq. (\ref{Eq.Torque_definition}) again gives zero except when $p_2 = 1$ and $p_1 = -1$. The expression for the torque becomes
\begin{eqnarray}\label{Eq.Torque_solution_p_1}
\tau &=& -\frac{\pi}{\mu_0}K_2 K_3 \textrm{sin}(\phi_0)
\end{eqnarray}
where the two constants $K_2$ and $K_3$ are given by
\begin{eqnarray}
K_2 &=& \Brem{}_{,1}\left(R_{\textrm{o},1}^{2} - R_{\textrm{i},1}^{2}\right)\\
K_3 &=& \Brem{}_{,2}\;\textrm{ln}\left(\frac{R_{\textrm{o},2}}{R_{\textrm{i},2}}\right)\nonumber\;.
\end{eqnarray}
Note that $K_2$ is identical to the constant $K_2$ in Eq. (\ref{Eq.Constants_torque_p_gt_1}) for $p_2 = 1$. We also see that Eq. (\ref{Eq.Torque_solution_p_1}) is in fact just $\mathbf{\tau} = \mathbf{m}\times\mathbf{B}$ 
for a dipole in a uniform field times the area of the magnet.

\subsection{Validating the expressions for the torque}
We have shown that there is only a torque between two Halbach cylinders if $p_1 = -p_2$ for $p_2 > 0$, with the torque being given by Eq. (\ref{Eq.Torque_solution}) for $p_2\neq 1$ and Eq. (\ref{Eq.Torque_solution_p_1}) 
for $p_2=1$.

\begin{figure}[!ht]
  \centering
   \includegraphics[width=\columnwidth]{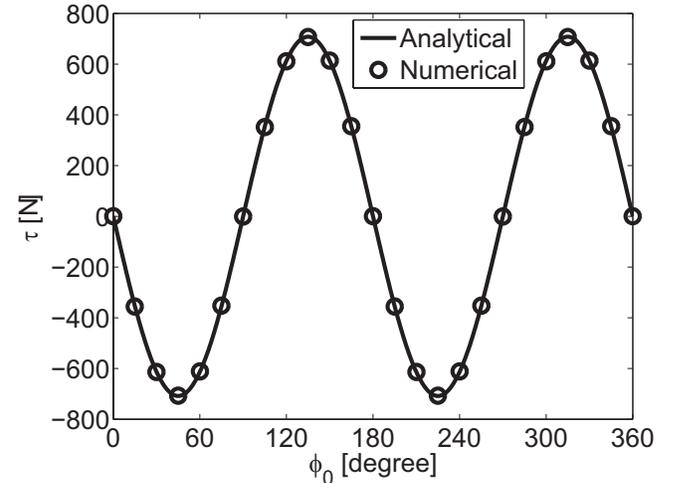}
      \caption{A numerical calculation of the torque per unit length between two concentric Halbach cylinders compared with the expression given in Eq. (\ref{Eq.Torque_solution}) for the physical properties given for Case
      1 in Table \ref{Table.Torque_cases}. The analytical expression is in excellent agreement with the numerical data. $\tau$ is per unit length as we consider a two dimensional system.}
       \label{Fig.Torque_compare_p_2}
\end{figure}

\begin{figure}[!ht]
  \centering
   \includegraphics[width=\columnwidth]{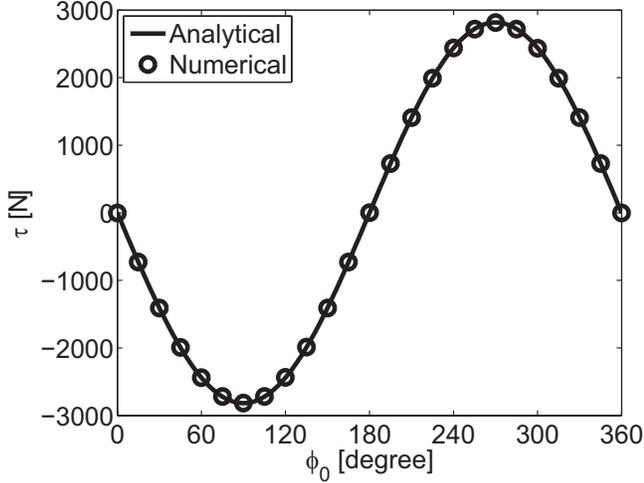}
      \caption{The torque per unit length given by Eq. (\ref{Eq.Torque_solution_p_1}) compared with a numerical calculation for the physical properties given for Case 2 in Table \ref{Table.Torque_cases}. As with the case
      for $p_2 \neq 1$, i.e. Fig. \ref{Fig.Torque_compare_p_2}, the analytical expression is in excellent agreement with the numerical data. $\tau$ is per unit length as we consider a two dimensional system.}
       \label{Fig.Torque_compare_p_1}
\end{figure}

To verify the expressions given in Eq. (\ref{Eq.Torque_solution}) and Eq. (\ref{Eq.Torque_solution_p_1}) we have computed the torque as a function of the angle of displacement, $\phi_0$, for the two cases given in Table 
\ref{Table.Torque_cases}, and compared this with a numerical calculation performed using Comsol. The results can be seen in Fig. \ref{Fig.Torque_compare_p_2} and \ref{Fig.Torque_compare_p_1}.

\begin{table}
\caption{\label{Table.Torque_cases}The parameters for the two cases shown in Fig. \ref{Fig.Torque_compare_p_2} and \ref{Fig.Torque_compare_p_1}.}
\begin{tabular}{ll|cccc}
                        & Magnet & $\Rin$ & $\Rou$ & $p$                  & $\Brem$ \\
                        &        & [mm]   & [mm]   &                      & [T]     \\ \hline
\multirow{2}{*}{Case 1:} & inner  &  5     &  15    &  -2                  & 1.4     \\
                        & outer  &  20    &  30    &   2                  & 1.4     \\ \hline
\multirow{2}{*}{Case 2:} & inner  &  10    &  35    &  -1                  & 1.4     \\
                        & outer  &  45    &  75    &   1                  & 1.4     \\
\end{tabular}
\end{table}

As can be seen from the figures the torque as given by Eq. (\ref{Eq.Torque_solution}) and Eq. (\ref{Eq.Torque_solution_p_1}) are in excellent agreement with the numerical results.

\section{Force and Torque for finite length cylinders}
The force and torque for finite length cylinders will be different than the analytical expressions derived above, because of flux leakage through the ends of the cylinder bore.

\begin{figure}[!ht]
  \centering
   \includegraphics[width=\columnwidth]{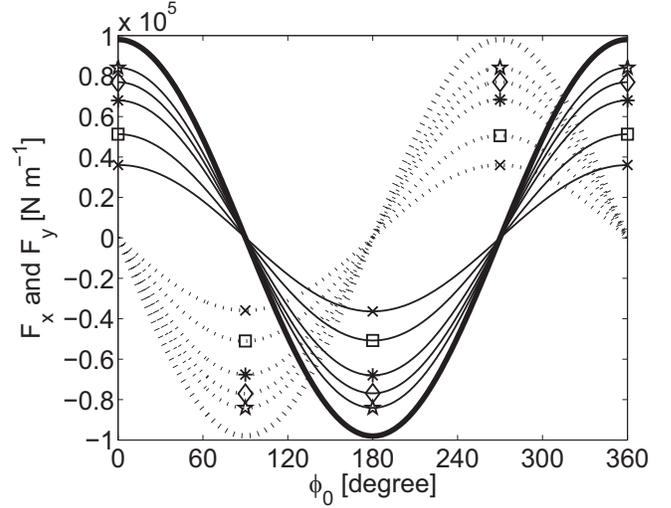}
      \caption{The two cartesian components of the force per unit length for a three dimensional system with dimensions as those given in Fig. \ref{Fig.Force_compare_pi_1_po_2}. The analytical expressions as well as the
      results of a three dimensional numerical simulation are shown.}
       \label{Fig.Force_3D_compare_pi_-1_po_2}
\end{figure}

To investigate the significance of this effect three dimensional numerical simulations of a finite length system corresponding to the system shown in Fig. \ref{Fig.Force_compare_pi_1_po_2} has been performed using Comsol. 
For this system the force has been calculated per unit length for different lengths. The results of these calculations are shown in Fig. \ref{Fig.Force_3D_compare_pi_-1_po_2}. From this figure it can be seen that as the 
length of the system is increased the force becomes better approximated by the analytical expression of Eq. (\ref{Eq.Force_solution}). A short system produces a lower force due to the leakage of flux through the ends of 
the cylinder. However, even for relatively short systems the two-dimensional results give the right order of magnitude and the correct angular dependence of the force.

\begin{figure}[!ht]
  \centering
   \includegraphics[width=\columnwidth]{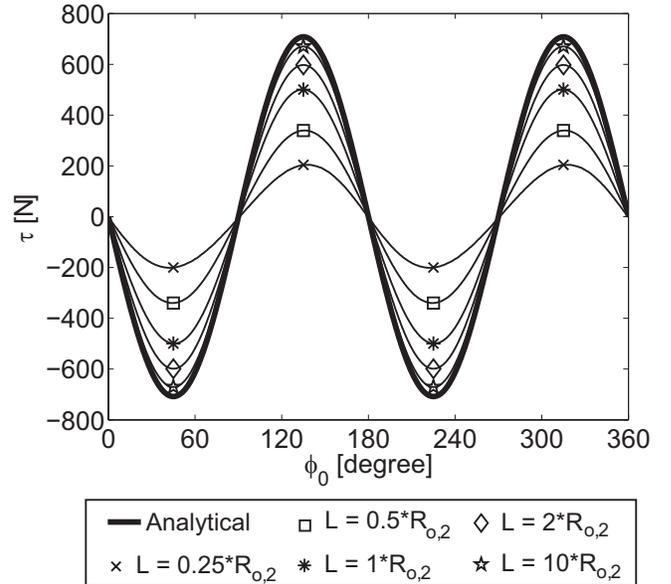}
      \caption{The torque per unit length for a three dimensional system with dimensions as those given as Case 1 in Table \ref{Table.Torque_cases}. The analytical expressions as well as the results of a three dimensional
      numerical simulation are shown.}
       \label{Fig.Torque_compare_3D_p_2}
\end{figure}

Similarly, the torque for a three dimensional system has been considered. Here the system given as Case 1 in Table \ref{Table.Torque_cases} was considered. Numerical simulations calculating the torque were performed, 
similar to the force calculations, and the results are shown in Fig. \ref{Fig.Torque_compare_3D_p_2}. The results are seen to be similar to Fig. \ref{Fig.Force_3D_compare_pi_-1_po_2}. The torque approaches the analytical 
expression as the length of the system is increased. As before the two dimensional results are still qualitatively correct.

Above we have considered cases where the two dimensional results predict a force ($p_{1} = 1-p_{2}$) or a torque ($p_{1} = -p_{2}$). However, for finite length systems a force or a torque can be present in other cases. One 
such case is given by Mhiochain et al. \cite{Mhiochain_1999} who report a maximum torque of $\approx 12$ Nm for a system where both magnets have $p = 1$, are segmented into 8 pieces and where the outer magnet has 
$R_{\textrm{i},2} = 52.5$ mm, $R_{\textrm{o},2} = 110$ mm, $L_{2} = 100$ mm and $\Brem{}_{,2} = 1.17$ T and the inner magnet has $R_{\textrm{i},1} = 47.5$ mm, $R_{\textrm{o},1} = 26$ mm, $L_{1} = 100$ mm and $\Brem{}_{,1} 
= 1.08$ T. This torque is produced mainly by the effect of finite length and to a lesser degree by segmentation. The torque produced by this system is $\approx 120$ N per unit length, which is significant compared to the 
expected analytical value of zero. The torque for finite length systems with $p_1\neq -p_2$ is, as noted above, a higher order effect. This makes it significantly smaller per unit length than for the corresponding system 
with $p_1 = -p_2$.

\ref{Fig.Torque_compare_3D_p_2} this system, which is designed to have a torque, produce a larger torque even though the system is much smaller.

The end effects due to a finite length of the system can be remedied by several different techniques. By covering the ends of the concentric cylinder with magnet blocks in the shape of an equipotential surface, all of the 
flux can be confined inside the Halbach cylinder \cite{Potenziani_1987}. Unfortunately this also blocks access to the cylinder bore. The homogeneity of the flux density can also be improved by shimming, i.e. placing small 
magnets or soft magnetic material to improve the homogeneity \cite{Abele_2006,Bjoerk_2008, Rowe_2008}. Finally by sloping the cylinder bore or by placing strategic cuts in the magnet the homogeneity can also be improved 
\cite{Hilton_2007}. However, especially the last two methods can lower the flux density in the bore significantly.

\section{Discussion and conclusion}
We have derived expressions for the magnetic vector potential, magnetic flux density and magnetic field for a two dimensional Halbach cylinder and compared these with numerical results.

The force between two concentric Halbach cylinders was calculated and it was found that the result depends on the integer $p$ in the expression for the remanence. If $p$ for the inner and outer magnet is termed $p_1$ and 
$p_2$ respectively it was shown that unless $p_1 = 1-p_2$ there is no force. The torque was also calculated for a similar system and it was shown that unless $p_1 = -p_2$ there is no torque. We compared the analytical 
expressions for the force and torque to numerical calculations and found an excellent agreement. Note that either there can be a force or a torque, but not both.

The derived expressions for the magnetic vector potential, flux density and field can be used to do e.g. quick parameter variation studies of Halbach cylinders, as they are much more simple than the corresponding three 
dimensional expressions.

An interesting use for the derived expressions for the magnetic flux density would be to derive expressions for the force between two concentric Halbach cylinders, where one of the cylinders has been slightly displaced. 
One could also consider the effect of segmentation of the Halbach cylinder, and of course the effect of a finite length in greater detail. Both effects will in general result in a nonzero force and torque for other choices 
of $p_1$ and $p_2$, but as shown these will in general be smaller than for the $p_1 = 1-p_2$ and $p_1 = -p_2$ cases.

It is also worth considering computing the force and torque for Halbach cylinders with $\mu_r \neq 1$. Here one would have to solve the complete magnetostatic problem of the two concentric Halbach cylinders to find the 
magnetic flux density in the gap between the cylinders.

\section*{Acknowledgements}
The authors would like to acknowledge the support of the Programme Commission on Energy and Environment (EnMi) (Contract No. 2104-06-0032) which is part of the Danish Council for Strategic Research.


\end{document}